\documentstyle[11pt]{article}

\newtheorem{lem}{Lemma}
\newtheorem{prop}{Proposition}
\newtheorem{cor}{Corollary}
\newtheorem{theo}{Theorem}
\newtheorem{deff}{Definition}

\def\ba{\begin{eqnarray}}
\def\ea{\end{eqnarray}}
\def\be{\begin{equation}}
\def\ee{\end{equation}}
\newfont{\msbm}{msbm10}
\newfont{\msbms}{msbm6}  
\newfont{\cmss}{cmss10}  
\def\U{{\cal U}}
\def\V{{\cal V}}
\def\W{{\cal W}}
\def\Q{{\cal Q}}

\def\S{{\cal S}}
\def\E{{\cal E}}

\def\H{{\cal H}}
\def\M{{\cal M}}
\def\M{{\cal M}}

\def\ss{\S(\R^d)}
\def\sdd{\S'(\R^d)}

\def\m{\mu}
\def\mum{\m^m}
\def\mum'{\m^{m'}}
\def\ms{\m_{\rho}}

\def\f{\phi}

\def\vf{\varphi}

\def\a{\alpha}
\def\b{\beta}
\def\s{\sigma}

\def\h2{${\rm h}(2)$}

\def\N{\hbox{\msbm N}}

\def\R{\hbox{\msbm R}}
\def\C{\hbox{\msbm C}}
\def\J{\hbox{\msbm J}}

\def\Ni{\hbox{\msbms N}}
\def\Ci{\hbox{\msbms C}}

\def\Ri{\hbox{\msbms R}}

\def\al{algebra}

\def\qu{quantization}

\def\rep{representation}
\def\reps{representations}

\def\trfs{transformations}

\def\bt{\begin{theo}}
\def\et{\end{theo}}

\begin{document}

\title{Canonical Quantization of the  Scalar Field: \\ The Measure Theoretic Perspective}
%%%%%%%%%%%%%%%%%%%%%%%%%%%%%%%%%%%%%%%%%%%%%%%%%%%%%%%%%%%%%%%%%%%%%%

%\author{J. M. Velhinho\thanks{Email:jvelhi@ubi.pt}\\
%Faculdade de Ci\^encias,  Universidade da Beira Interior, \\ 
%R. Marqu\^es D'\'Avila e Bolama, 6201-001 Covilh\~a, Portugal}
\author{Jos\'e Velhinho}

\date{{Faculdade de Ci\^encias, Universidade da Beira 
Interior\\R. Marqu\^es D'\'Avila e Bolama,
6201-001 Covilh\~a, Portugal}\\{jvelhi@ubi.pt}}

\maketitle

\begin{abstract}
\noindent
This review is devoted to measure theoretical methods in the canonical quantization
of scalar field theories. We present in some detail the canonical quantization of the free scalar
field. We study the measures associated with the free fields
and present two  characterizations of the support of these measures.
The first characterization concerns local properties of the quantum fields,
whereas for the second one
we introduce a sequence of variables that test the field
behaviour at large distances, thus allowing to distinguish between the 
typical quantum fields associated with different values of the mass.
\end{abstract}

\section{Introduction}
\label{intro}

The phase space of the classical scalar field is a  linear space, and  therefore constitutes
an infinite dimensional analogue of the usual phase space for classical dynamics of a finite number of particles, namely the cotangent bundle $T^*\R^n$.
The usual finite dimensional  Heisenberg kinematical algebra admits a natural generalization
in this infinite dimensional context: the kinematical variables are conveniently labelled  by
smooth test functions, belonging to the  real Schwartz  space $\S(\R^d)$. The Heisenberg  group and the Weyl 
relations admit suitable generalizations as well, and therefore the problem 
of the canonical quantization of kinematical observables in scalar field theory is, {\it a priori}, well defined: following Weyl, Gelfand and Segal,
one should look for representations of the Weyl relations  \cite{BSZ,RS2}. 
However, in contrast to what happens  in finite dimensions, the Weyl relations in field theory
admit  nonequivalent representations, i.e., the quantization of the kinematics is not unique. 
Note that this is definitely not due only to the existence of pathological representations: examples of physically  relevant nonequivalent
representations are those associated with free fields with different masses  \cite{RS2}. 
Moreover, the quantization of the kinematics of theories with interactions is not equivalent
to the quantization of the kinematics of free theories \cite{RS3,H}. 
So, the dynamics plays a crucial role in the selection/construction 
of a quantum representation adequate to a given classical field model, already at the kinematical level.  
It is also important the issue of the unitary representation, at the quantum level, of natural symmetries
of the classical model, e.g. the 
Poincar\'e group.
Note that  a unitary representation of the  Poincar\'e group includes a quantization
of the dynamics, given  by the representation of the subgroup of time translations.
This is essentially the problem of quantization of field theories: given a certain
classical model, to construct a quantum representation with the appropriate invariance properties,
 therefore allowing a consistent quantization of the dynamics and of relevant symmetry groups.

The theory of  \reps\ of the  Weyl relations can be seen as a problem in measure theory 
in infinite dimensional linear spaces.
Just as in finite dimensions, it would be natural to look for
 \reps\ in spaces of square integrable functions with respect to some measure on the classical
 configuration space, which is a space of functions
in $\R^d$. It turns out, however, that potentially interesting measures on those spaces
fail to satisfy the crucial property of 
 $\s$-additivity, which, in particular, implies that one cannot obtain Hilbert spaces out of the measures.
To obtain  $L^2$ spaces it is necessary to extend the classical configuration space
to spaces of distributions.
Spaces of distributions are indeed the natural ``home'' to interesting measures in field theory. 
In order to obtain  \reps\ of the Weyl relations it is
 sufficient to consider the space of distributions \ $\S'(\R^d)$, 
dual to the  Schwartz space of test functions $\S(\R^d)$. In fact, one can show that there is a one-to-one
correspondence between (cyclic) 
\reps\ and certain  classes of measures in $\S'(\R^d)$ \cite{GV}. 
These measures have the  property of being 
quasi-invariant with respect to the action of  $\S(\R^d)$ as translations in
 $\S'(\R^d)$, which   essentially means that the translation of such a measure  by an element of 
  $\S(\R^d)$ produces a measure  supported on the same subset of  $\S'(\R^d)$. 
Measures $\m$ of this type define 
\reps\ of the  Weyl relations in the  Hilbert space 
 of square integrable functions $L^2\bigl(\S'(\R^d),\m\bigr)$ \cite{GV}. The distribution space 
$\S'(\R^d)$ can therefore be seen as the  ``universal quantum configuration space'' for
the real scalar field. 
%%%
%These measures, however, are  not quasi-invariant with respect 
%relativamente a todas as transla\coes . De facto, n\~ao existem medidas em 
%$\S'(\R^d)$ quasi-invariantes para todas as transla\coes\ \cite{GV,Ya}. 
%This fact,
%simultaneamente t\'\i pico e exclusivo de dimens\~ao infinita \cite{GV,Ya},
%est\'a na origem da exist\^encia de \reps\ irredut\'\i veis n\~ao-equivalentes
%das rela\coes\ de Weyl \cite{BSZ,RS2}.
%%%
\smallskip

In the present review we discuss in detail the canonical quantization of the free massive scalar field,
following \cite{BSZ,RS2,GV,GJ}, including also
an analysis of the support of the corresponding measures. This latter
study follows closely \cite{MTV} and \cite{ZeS}, which in turn were partly inspired in \cite{RR,CL,Ri}. 
In Section \ref{gauss} we have collected a minimal amount of relevant notions
and results concerning Gaussian \reps\ of the Weyl relations.
Section \ref{section1} deals with the classical 
dynamics in the Hamiltonean formalism: since the equations of motion are 
linear, the solution is given by a one-parameter group of linear sympletic 
transformations. In section \ref{section2} we discuss the unitary 
implementation in the quantum theory of linear sympletomorphisms, in the 
context of Gaussian \reps\ of the Weyl relations. In section \ref{section3} 
we present the Gaussian measure corresponding to the free quantum field. 
We will see that for each value of the mass there is a corresponding
Gaussian measure allowing a physically consistent quantization of the dynamics.
Quantum dynamics is given by an one-parameter group of unitary 
transformations in perfect correspondence with the classical situation. One 
can show that the quantum 
Hamiltonean is a positive operator and that there is a  unique vacuum. 
Moreover, it can  be shown 
that these 
conditions determine an unique quantum \rep\ \cite{BSZ}. In section 
\ref{section4} we show that the free field measure is invariant and ergodic 
with respect to the action of the Euclidean group on $\R^d$. This result gives us an 
unitary \rep\ of the Euclidean group on the quantum Hilbert space and 
shows that the 
vacuum is the only invariant state. Since this is true for every value of the 
mass, it also shows that the \reps\ of the Weyl relations corresponding to 
two different values of the mass are not unitarily equivalent.
In section \ref{section5} we discuss briefly the relativistic invariance 
properties of the free field quantization.
The last two sections concern properties of the support of the free field measures.
In Section \ref{s2} local properties of the support are discussed.
In section \ref{clivre2} we analyse the long range behaviour instead, which will allow us  to
distinguish between the supports of the measures  associated with different values of the
mass.

%%%%%%%%%%%%%%%%%%%%%%%%%%%%%  Section  %%%%%%%%%%%%%%%%%%%%%%%%%%%%%%%%%
\section{Gaussian \reps\ of the Weyl relations}
\label{gauss}
The canonical quantization of field theories involves the introduction
of a convenient measure in an infinite dimensional space.
In the case of the real scalar field the appropriate measure space is the 
dual of the Schwartz space. Measures in this space allow the construction
of representations of the Weyl relations which are of the  Schr\"odinger type.
In particular, the measure associated with the free field is a Gaussian measure.

\subsection{Gaussian measures on $\sdd$}
\label{appendixA}
%%%%%%%%%%%%%%%%%%%%%%%%%%%%%%%%%%%%%%%%%%%%%%%%%%%%%%%%%%%%%%%%%%%%%

Let $\sdd$ be the topological dual of the real Schwartz space $\ss$,
with respect to the the nuclear topology. We will consider $\sdd$ as a measurable space,
the $\s$-algebra of measurable sets being the smallest $\s$-algebra 
such that all the maps $\phi\mapsto\phi(f)$, $\phi\in\sdd$,
$f\in\ss$, are measurable. This $\s$-algebra coincides with the
Borel $\s$-algebra associated with the strong topology on $\sdd$.
%%%%%%%%%%%%%%%%%%%%%%%%%%%%%
\begin{deff}
\label{appdef1}
The Fourier transform of a measure $\m$ on $\sdd$ is the function
$\chi:\ss\to\C$ defined by
%%%%%%%%%%%%
\be
\label{app1}
\chi(f)=\int_{\S'(\Ri^d)} e^{i\phi(f)}\,d\mu(\phi)\,.
\ee
%%%%%%%%%%%
\end{deff}
%%%%%%%%%%%%%%%%%%%%%%%%%%%%%%%
%%%%%%%%%%%%%%%%%%%%%%%%%%%%%%%%
\begin{deff}
\label{appdef2}
A complex function $\chi$ on a linear space $E$ is said to be of
positive type if
%%%%%%%%%%%%%%%%
\be
\label{app2}
\sum^m_{k,l=1}c_k\,\bar c_l\,\chi(\xi_k-\xi_l)\geq 0,\ \ 
\forall m\in\N,\ c_1,\ldots,c_m\in\C,\ \xi_1,\ldots,\xi_m\in E\,.
\ee
%%%%%%%%%%%%%%
\end{deff}
%%%%%%%%%%%%%%%%%%%%%%%%%%%%%%%%%%
%%%%%%%%%%%%%%%%%%%%%%%%%%%%%%%%%%
\begin{theo}[Bochner-Minlos]
\label{appteo1}
The Fourier transform of a measure on $\sdd$ is a positive type function,
continuous with respect to the nuclear topology. Conversely, a continuous function 
of positive type on $\ss$ is the Fourier transform of an uniquely defined
measure on $\sdd$.
\end{theo}
%%%%%%%%%%%%%%%%%%%%%%%%%%%%%%%%%%
%%%%%%%%%%%%%%%%%%%%%%%%%%%%%%%%%%
\begin{deff}
\label{appdef3}
Let $(\,,)$ be a continuous inner product on $\ss$. The Gaussian measure
on $\sdd$ of covariance $(\,,)$ is the measure whose Fourier transform
is $\chi(f)=e^{-(f,f)/2}$, $f\in\ss$.
\end{deff}
%%%%%%%%%%%%%%%%%%%%%%%%%%%%%%%%%
Note that given a Gaussian measure $\mu$ of covariance $(\,,)$, every
element of the real Hilbert space $\H$, the completion of $\ss$
with respect to $(\,,)$, still defines an element of $L^1\bigl(\sdd,\mu\bigr)$,
generalizing $\f\mapsto\f(f)$.
This follows (see e.g.~\cite[I.1, I.2]{S}) from the obvious fact that the 
Fourier transform is
continuous with respect to the $(\,,)$ norm.

We are particularly interested in the case where the covariance is defined by
certain types of linear operators on $\ss$.
%%%%%%%%%%%%%%%%%%%%%%%%%%%%%%%%%%%%
\begin{deff}
\label{appdef4}
We will say that a linear operator $C:\ss\to\ss$ is a covariance operator if:
\begin{itemize} 
\item[{\rm (}i\/{\rm )}] $C$ is a homeomorphism of $\ss$ with respect to the
nuclear topology;
\item[{\rm (}ii\/{\rm )}] $C$ is bounded, self-adjoint and positive
on $L^2(\R^d)$;
\item[{\rm (}iii\/{\rm )}] $C^{-1}$, seen as a densely defined operator on 
$L^2(\R^d)$, is (essentially) self-adjoint and positive.
\end{itemize}
\end{deff}
%%%%%%%%%%%%%%%%%%%%%%%%%%%%%%%%%%%%
It is clear that the bilinear form
%%%%%%%%%%%%
\be
\label{app3}
\langle f,g\rangle_C:=\langle f,Cg\rangle,\ \ f,g\in\ss
\ee
%%%%%%%%%%%%
defines an inner product, for any covariance operator, where $\langle\,, \rangle$ denotes the $L^2(\R^d)$
inner product, i.e.
\be
\langle f,g\rangle:=\int_{\Ri^d}fg\,d^dx.
\ee
A covariance operator
$C$ defines thus a Gaussian measure, and we will say also that $C$ is
the covariance of the measure.

\subsection{Representations of the Weyl relations}
\label{appendixB}

In the canonical quantization of real scalar field theories in $d+1$
dimensions one looks for unitary \reps\ of the Weyl relations
%%%%%%%%%%%%%%%
$$
\V(g)\,\U(f)=e^{i\langle f,g\rangle}\,\U(f)\V(g)\,,
$$
%%%%%%%%%%%%%%%%
where $f$ and $g$ belong to $\ss$. By \rep\ of the above relations
it is meant a pair $(\U,\V)$ of strongly continuous unitary \reps\
(on the same Hilbert space) of the commutative nuclear group $\ss$.
It is also required that the combined action of $\U$ and $\V$ be
irreducible. A given \rep\ on a Hilbert space $\H$ is said to be cyclic
if there is $\theta\in\H$ such that the linear space of $\bigl\{\U(f)\theta,\ 
f\in\ss\bigr\}$ is dense in $\H$. We will consider only cyclic \reps .

It is a well established fact \cite{GV} that cyclic \reps\ are in one to
one correspondence with quasi-invariant measures on $\sdd$. 
Consider first the action of $\ss$ as translations of $\sdd$:
%%%%%%%%%%%%%%%%%
\be
\label{app4}
\sdd\ni\phi\mapsto\phi+f,\ \ f\in\ss
\ee
%%%%%%%%%%%%%%%%%
where $\phi+f$ is defined by $(\phi+f)(g)=\phi(g)+\langle f,g\rangle$,
$\forall g\in\ss$. A measure $\mu$ on $\sdd$ is quasi-invariant
(with respect to the above action) if the translated measures $\mu_f(\phi):=\mu(\phi-f)$
are mutually absolutely continuous with respect to $\mu$, $\forall f\in\ss$.
(Note that there are no quasi-invariant measures with respect to all the translations
on $\sdd$. This fact is ultimately responsible for the existence of
nonequivalent \reps\ of the Weyl relations, in contrast with the
corresponding situation in finite dimensions.)

Given a quasi-invariant measure $\mu$, one defines a cyclic \rep\ of
the Weyl relations on $L^2\bigl(\sdd,\mu\bigr)$ by the following actions of
$\U$ and $\V$:
%%%%%%%%%%%%%%%
\begin{eqnarray}
\label{app5}
\bigl(\U(f)\psi\bigr)(\phi) & = & e^{-i\phi(f)}\psi(\phi)\,, \\
\label{app6}
\bigl(\V(g)\psi\bigr)(\phi) & = & \left({d\mu(\phi-g)\over d\mu(\phi)}
\right)^{1/2}\psi(\phi-g)\,.
\end{eqnarray}
%%%%%%%%%%%%%%%%
The \rep\ $\V$ is unitary precisely because the measure is quasi-invariant,
allowing the existence of the Radon-Nikodym derivative.

The following proposition gives necessary and sufficient conditions for
the equivalence of cyclic \reps . We introduce the Weyl operators
$\W(f,g):=e^{i\langle f,g\rangle/2}\U(f)\V(g)$.
%%%%%%%%%%%%%%%%%%%%%%%%%%%%%%
\begin{prop}
\label{appprop1}
Two cyclic \reps\ $(\H,\W)$ and $(\H',\W')$ with cyclic vectors $\theta$
and $\theta'$, respectively, satisfy
%%%%%%%%%%%%%%%%
\be
\label{app7}
\bigl\langle \theta',\W'(f,g)\theta'\bigr\rangle=
\bigl\langle \theta,\W(f,g)\theta\bigr\rangle,\ \ \forall f,g
\ee
%%%%%%%%%%%%%
if and only if there is an unitary operator $T:\H'\to\H$ 
such that $T\,\W'(f,g)\,T^{-1}=\W(f,g)$ and $T\theta'=\theta$.
\end{prop}
%%%%%%%%%%%%%%%%%%%%%%%%%%%%%%%%
The \reps\ (\ref{app5}, \ref{app6}) are irreducible if and only if
the measure is ergodic with respect to the action (\ref{app4}) of $\ss$
\cite{BSZ}. Moreover, one can show that two ergodic measures give rise to
unitarily equivalent \reps\ if and only if the two measures are
mutually absolutely continuous. 
Well known examples of quasi-invariant and ergodic measures on $\sdd$ are
provided by Gaussian measures. We will say that the corresponding \reps\
of the Weyl relations are Gaussian \reps .

Let then $C$ be a covariance operator on $\ss$ and $\mu$ the corresponding
measure on $\sdd$. The Radon-Nikodym derivative in (\ref{app6}) is
easy to evaluate and one thus have the following irreducible \rep\
of the Weyl relations defined by the covariance $C$:
%%%%%%%%%%%%%%%
\begin{eqnarray}
\label{app8}
\bigl(\U(f)\psi\bigr)(\phi) & = & e^{-i\phi(f)}\psi(\phi)\,, \\
\label{app9}
\bigl(\V(g)\psi\bigr)(\phi) & = & e^{-\langle g,C^{-1}g\rangle/4}
\,e^{\phi(C^{-1}g)/2}\,\psi(\phi-g)\,.
\end{eqnarray}
%%%%%%%%%%%%%%%%
One can easily evaluate the expectation values of the Weyl operators on
the cyclic vector:
%%%%%%%%%%%%%%%%
\be
\label{app10}
\bigl\langle 1,\W(f,g)1\bigr\rangle=e^{-\left\langle\!\left\langle (f,g),
(f,g)\right\rangle\!\right\rangle_C/4}
\ee
%%%%%%%%%%%%%%%%
where $\langle\!\langle\,,\rangle\!\rangle_C$ is the inner product on
$\ss\oplus\ss$:
%%%%%%%%%%%%%%%%
\be
\label{app11}
\bigl\langle\!\bigl\langle (f,g),(f',g')\bigr\rangle\!\bigr\rangle_C:=
\bigl\langle f,(2C)f\bigr\rangle+\bigl\langle g,(2C)^{-1}g\bigr\rangle\,.
\ee
%%%%%%%%%%%%%%%%%
In the case of Gaussian \reps\ the natural topology on the test functions 
space is determined by the inner product 
$\langle\!\langle\,,\rangle\!\rangle_C$. It is not  difficult to show 
that $\U$ (\ref{app8}) and $\V$ (\ref{app9}) are in fact continuous with respect to the
inner products $\langle\cdot\,,C\,\cdot\rangle$ and 
$\langle\cdot\,,C^{-1}\,\cdot\rangle$, respectively. 
Therefore, the Weyl operators
are well defined for all $f,g\in\H_C\oplus\H_{C^{-1}}$, the real completion
of $\ss\oplus\ss$ with respect to $\langle\!\langle\,,\rangle\!\rangle_C$.
Equation (\ref{app10}) for the expectation values still hold.

\section{Classical free field dynamics}
\label{section1}
%%%%%%%%%%%%%%%%%%%%%%%%%%%%%%%%%%%%%%%%%%%%%%%%%%%%%%%%%%%%%%%%%%%%%%%%%

Let us consider the free scalar field of mass $m$ in $d+1$ dimensions, whose 
dynamics is given by the Klein-Gordon equation
%%%%%%%%%%%%%%%%%%%%%%%%%%%%
\be
\label{II1.2.1}
\Box \varphi +m^2\varphi =0\,,
\ee
%%%%%%%%%%%%%%%%%%%%%%%%%%%%%%
where $\Box $  is the d'Alembert operator in $d+1$ dimensions, 
$\Box :={\partial ^2 \over \partial t^2}-\sum _{i=1}^d
{\partial ^2 \over \partial x_i^2}$. Considering the momenta 
$\pi:=\partial\varphi/\partial t$, one can rewrite (\ref{II1.2.1}) as
a first order system
%%%%%%%%%%%%%%%%%%%%%%%%%%%%%%%%%%%
\ba
\label{II1.2.4}
{\partial \varphi \over \partial t} & = & \pi \\
\label{II1.2.5}
{\partial \pi \over \partial t} & = & -(m^2-\Delta )\varphi \,,
\ea
%%%%%%%%%%%%%%%%%%%%%%%%%%%%%%%%%%%
where $\Delta:=\sum_{i=1}^d{\partial^2\over\partial x_i^2}$. The evolution
equations (\ref{II1.2.4}, \ref{II1.2.5}) can be written in Hamiltonean form
%%%%%%%%%%%%%%%%%%%%%%%%%%%%%%%
\ba
\label{II1.1.5}
{\partial \varphi \over \partial t} & = & \{\varphi ,H\}\\
\label{II1.1.6}
{\partial \pi \over \partial t} & = & \{\pi ,H\}\,,
\ea
%%%%%%%%%%%%%%%%%%%%%%%%%%%%%%%
with Hamiltonean function given by
%%%%%%%%%%%%%%%%%%%%%%%%%%%%%%%
\be
\label{II1.2.3}
H={\textstyle {1 \over 2}}\int d^dx \bigl(\pi ^2+
\varphi (m^2-\Delta )\varphi \bigr)
\ee
%%%%%%%%%%%%%%%%%%%%%%%%%%%%%%%%%%
and Poisson bracket defined in the usual way, involving functional derivatives.

The operator $m^2-\Delta$ plays an important role in what follows. The
following proposition collects some properties of this operator
\cite{RS2,GJ,Ru}.
%%%%%%%%%%%%%%%%%%%%%%%%%%%%%%%%%
\begin{prop}
\label{propII1.2.1}\begin{itemize} \item[ ]
\item[{\rm (}i\/{\rm )}] The differential operator $m^2-\Delta$ is a 
continuous linear operator in the real Schwartz space $\ss$, with respect to the
nuclear topology. The same is true  for the inverse operator
$(m^2-\Delta)^{-1}$. When considered as an operator on the complex
Schwartz space $\S^{\Ci}(\R^d)$, $m^2-\Delta$ defines a self-adjoint 
positive operator on $L^2(\R^d)$. The operators 
$(m^2-\Delta)^{l/2^j}$, $l,j \in \N $, are densely defined in $L^2(\R^d)$ 
and enjoy the same above mentioned properties.

\item[{\rm (}ii\/{\rm )}] The inverse operator $(m^2-\Delta)^{-1}$ 
is bounded, self-adjoint and positive on $L^2(\R ^d)$. The operators
$(m^2-\Delta)^{-l/2^j}\!,\ l,j \in \N $, are well defined and enjoy the same 
properties.

\item[{\rm (}iii\/{\rm )}] The operators $(m^2-\Delta)^{\pm l/2^j}$, $l,j
\in \N $, are strictly positive on $L^2(\R^d)$, i.e., $\langle\psi,
(m^2-\Delta)^{\pm l/2^j}\psi\rangle=0$ for $\psi$ in the domain 
implies $\psi=0$.
\end{itemize}
\end{prop}
%%%%%%%%%%%%%%%%%%%%%%%%%%%%%%%%%%%%%%%%%%%%%%%%%%%%%%%%%%%%%%%%%%%%%%
The evolution equations (\ref{II1.2.4}, \ref{II1.2.5}) are easily solved,
in an appropriate Hilbert space. We follow here \cite[X.13]{RS2} and
\cite[6.2]{BSZ}. Formally, the fundamental solution of
(\ref{II1.2.4}, \ref{II1.2.5}) is given by the kernel of the evolution 
operator
%%%%%%%%%%%%%%
\be
\label{livre com massa 1}
\Theta_m (t)=\exp\biggl(\biggl(\begin{array}{cc} 0 & 1 \\ -(m^2-\Delta) & 0
\end{array}\biggr)\,t\biggr)
\ee
%%%%%%%%%%%%%%%
that one obtains by exponentiation of the operator 
$\biggl(\begin{array}{cc} 0 & 1 \\ -(m^2-\Delta) & 0 \end{array}\biggr)$ 
in the system (\ref{II1.2.4}, \ref{II1.2.5}). 
The rigorous treatment of this result, 
however, requires an appropriate Hilbert structure, that we now describe.
Let $\widetilde\H^{\pm}_m$ be the Hilbert spaces obtained by completion
of $\S^{\Ci}(\R^d)$ with respect to the inner products
%%%%%%%%%%%%
\be
\label{livre com massa 2}
(f,g)\mapsto\bigl\langle f,(m^2-\Delta)^{\pm 1/2}g\bigr\rangle\,.
\ee
%%%%%%%%%%%%
The operator 
%%%%%%%%%%
$$
i\biggl(\begin{array}{cc} 0 & 1 \\ -(m^2-\Delta) & 0 
\end{array}\biggr)
$$ 
%%%%%%%%%%%%
is densely defined in $\widetilde\H^+_m\oplus\widetilde\H^-_m$ and is
self-adjoint \cite[X.13]{RS2}. Then, (\ref{livre com massa 1}) gives us
a well defined unitary \rep\ of $\R$ on 
$\widetilde\H^+_m\oplus\widetilde\H^-_m$, which is strongly continuous.
It can be shown \cite[X.13]{RS2} that the operators $\Theta_m (t)$ 
(\ref{livre com massa 1}) preserve the real subspaces of $\widetilde\H^+_m\oplus\widetilde\H^-_m$.
Thus, the operators $\Theta_m (t)$ are orthogonal on the real Hilbert
space $\H^+_m\oplus\H^-_m$, where $\H^{\pm}_m$ is the completion of the
real Schwartz space $\ss$ with respect to (\ref{livre com massa 2}). The dynamics
of the free field of mass $m$ is therefore well defined on the space
$\H^+_m\oplus\H^-_m$, which can then be taken as the appropriate classical 
phase space
for this system. The sympletic form in this linear space is given by
%%%%%%%%%%%%%
\be
\label{n1}
\Omega\bigl((\vf_1,\pi_1),(\vf_2,\pi_2)\bigr)=\langle\vf_1,\pi_2\rangle -
\langle\vf_2,\pi_1\rangle \,,
\ee
%%%%%%%%%%%%%%
where the map
%%%%%%%%%%%%%
\be
\label{livre com massa 3}
\H^+_m\oplus\H^-_m\ni (\vf,\pi)\mapsto\langle\vf,\pi\rangle=
\int_{\Ri^d} \vf\,\pi\,d^dx
\ee
%%%%%%%%%%
is well defined and continuous.

Let $\langle\!\langle\,,\rangle\!\rangle_{-m}$ be the $\H^+_m\oplus\H^-_m$
inner product. The form $\Omega$ (\ref{n1}) can be written as
%%%%%%%%%%%%%%%
\ba
\label{livre com massa 4}
\Omega\bigl((\vf_1,\pi_1),(\vf_2,\pi_2)\bigr)= \!\!\!\!
& \biggl\langle\!\!\!\biggl\langle & \!\!\!\!(\vf_1,\pi_1),\biggl(
\begin{array}{cc} (m^2-\Delta)^{-1/2} & 0 \\ 0 & (m^2-\Delta)^{-1/2} 
\end{array}\biggr)\cdot \nonumber \\
& \cdot & \!\!\!\!\!\biggl(\begin{array}{cc} 0 & 1 \\ -(m^2-\Delta) & 0 
\end{array}\biggr)(\vf_2,\pi_2)\biggr\rangle\!\!\!\biggr\rangle_{-m}\,.
\ea
%%%%%%%%%%%%%%%
It is then clear that the group $\Theta_m (t)$ preserves the sympletic form,
given that $\Theta_m (t)$ is orthogonal in $\H^+_m\oplus\H^-_m$, $\forall t$, 
and commutes with
%%%%%%%%%
$$
\biggl(\begin{array}{cc} (m^2-\Delta)^{-1/2} & 0 \\ 0 & 
(m^2-\Delta)^{-1/2}\end{array}\biggr)
$$
%%%%%%%%%
and with
%%%%%%%%%%
$$
\biggl(\begin{array}{cc} 0 & 
1 \\ -(m^2-\Delta) & 0 \end{array}\biggr)\,.
$$
%%%%%%%%%
Let us consider the evolution of the kinematical observables $F_{c,(f,g)}$:
%%%%%%%%%%%%%%
\be
\label{n2}
F_{c,(f,g)}(\vf,\pi):=c+\langle f,\vf\rangle+\langle g,\pi\rangle\,,
\ee
%%%%%%%%%%%%%%%%
where $c\in\R$ and $f,g\in\ss$. For each $t\in\R$, let $F_{c,(f,g)}^t$ be the
pull-back of $F_{c,(f,g)}$:
%%%%%%%%%%%%
\be
\label{livre com massa 5}
F_{c,(f,g)}^t(\vf,\pi):=F_{c,(f,g)}\bigl(\Theta_m (t)(\vf,\pi)\bigr)\,.
\ee
%%%%%%%%%%%%
It is clear that $F_{c,(f,g)}^t=F_{c,S_m(t)(f,g)},\ \forall t\in\R$, where
%%%%%%%%%%%%%%%%
\be
\label{livre com massa 6}
S_m(t):=\exp\biggl(\biggl(\begin{array}{cc} 0 & -(m^2-\Delta) \\ 1 & 0 
\end{array}\biggr)\,t\biggr)
\ee
%%%%%%%%%%%%%%%
is an orthogonal operator on the real Hilbert space $\H^-_m\oplus\H^+_m$.
The linear space $\R\oplus(\H^-_m\oplus\H^+_m)$ of test functions is
a Lie algebra under the Lie bracket
%%%%%%%%%%%%
\be
\label{n3}
\Bigl[\bigl(c,(f,g)\bigr),\bigl(c',(f',g')\bigr)\Bigr]=
\bigl(\langle f,g'\rangle-\langle g,f'\rangle,(0,0)\bigr)\,.
\ee
%%%%%%%%%%%%%
This \al\ is obviously isomorphic 
to the algebra of kinematical functions (\ref{n2}) under the Poisson bracket.
The above result shows that the dynamics is implemented on the
kinematical \al\ as a group of linear automorphisms. The group $S_m(t)$ 
(\ref{livre com massa 6}) is an orthogonal \rep\ of $\R$ on
$\H^-_m\oplus\H^+_m$.

%%%%%%%%%%%%%%%%%%%%%%%%%%%%%  Section  %%%%%%%%%%%%%%%%%%%%%%%%%%%%%%%%%
\section{Unitary implementation of linear sympletomorphisms}
\label{section2}
%%%%%%%%%%%%%%%%%%%%%%%%%%%%%%%%%%%%%%%%%%%%%%%%%%%%%%%%%%%%%%%%%%%%%%%%%

In this section we discuss the question of unitary implementability of
linear canonical transformations, in the context of Gaussian \reps\ of
the Weyl relations for the real scalar field. 
We follow \cite{BSZ}, although considering real Hilbert 
spaces instead of complex ones, in the description of the classical
phase space and kinematical \al .

Let $A$ be a linear transformation on $\S(\R^d)\oplus\S(\R^d)$, continuous
in the nuclear topology and with continuous inverse. Like in finite dimensions,
we will say that $A$ is sympletic if $A$ preserves the sympletic matrix
$\J=\left(\begin{array}{cc} 0 & {\bf 1} \\ -{\bf 1} & 0 \end{array}\right)$
in $\S(\R^d)\oplus\S(\R^d)$. We just saw in the last section that the 
classical evolution of the free field of mass $m$ is determined by a
one-parameter group of linear sympletic transformations $S_m(t)$ 
(\ref{livre com massa 6}). To quantize the system, one needs a \rep\ of the 
Weyl relations allowing the unitary implementation of this group. In more
precise terms, one looks for a \rep\ $(\H_m,\W_m)$ of the
Weyl relations by Weyl operators $\W_m(f,g)$, $f,g\in\ss$, on a (complex)
Hilbert space $\H_m$ such that there exists a group of unitary transformations
$T_m(t):\H_m\to\H_m$, $t\in\R$, satisfying
%%%%%%%%%%%%%%
\be
\label{iusl0}
T_m(t)\W_m(f,g)T_m(t)^{-1}=\W_m\bigl(S_m(t)(f,g)\bigr)\,.
\ee
%%%%%%%%%%%%%%%%
For the free field, (\ref{iusl0}) is
satisfied by a Gaussian \rep . In fact, the (equivalence class of the) \rep\
$(\H_m,\W_m)$ is uniquely determined by the dynamics, i.e., by 
(\ref{iusl0}) and by the natural conditions of positivity of the quantum 
Hamiltonean and unicity of the vacuum. This result is based on the theorem 
below \cite{BSZ}. Before presenting the theorem, let us illustrate the
non-triviality of the \qu\ process for sympletomorphisms (in general for
observables outside the kinematical \al ).

Let $(\H,\W)$ be a continuous and irreducible \rep\ of the Weyl relations and
$A$ a linear sympletomorphism. We can define a new continuous and
irreducible \rep\ $(\H,\W_A)$ on the same Hilbertspace $\H$, by
%%%%%%%%%%%%%%
\be
\label{iusl1}
\W_A(f,g):=\W\bigl(A(f,g)\bigr)\,.
\ee
%%%%%%%%%%%%%
In finite dimensions the Stone-von Neumann theorem shows that $\W$ and $\W_A$
are unitarily equivalent, and therefore guarantees the existence of an
unitary operator $U(A)$  corresponding to the \qu\ of $A$. In infinite
dimensions, however, the \reps\ $\W$ and $\W_A$ are not necessarily equivalent 
and therefore the \qu\ of a given canonical transformation $A$ does not
necessarily exist for an arbitrary \rep\ $\W$. Gaussian \reps\ give us good
examples of this fact, as we now show. Let $A_{\s}$ be the sympletic
transformation given by
%%%%%%%%%%%%%%
$$
(f,g)\mapsto (\s^{1/2}f,\s^{-1/2}g),\ \ \s\in\R^+,\ \s\neq 1\,.
$$
%%%%%%%%%%%%%%
For convenience of notation, let us denote by $\W^{C}$ the Gaussian \rep\
on the Hilbert space $L^2\bigl(\sdd,\mu\bigr)$, defined by the Gaussian
measure $\mu$ of covariance $C:\ss\to\ss$. 
Let us consider the new
\rep 
%%%%%%%%%%%%%%%%%%%
$$
\W_{A_{\s}}^{C}(f,g):=\W^{C}(\s^{1/2}f,\s^{-1/2}g)\,.
$$
%%%%%%%%%%%%%%%%%
For every $f,g\in\ss$ one has
%%%%%%%%%%%%%%%
$$
\bigl\langle 1,\W_{A_{\s}}^{C}(f,g)1\bigr\rangle=
\bigl\langle 1,\W^{\s C}(f,g)1\bigr\rangle
$$
%%%%%%%%%%%%%%%%
and therefore the two \reps\ $\W_{A_{\s}}^{C}$ and $\W^{\s C}$ are
unitarily equivalent (see section \ref{appendixB}). 
On the other hand, the \reps\ $\W^{C}$ and 
$\W^{\s C}$ are not equivalent, given that the corresponding measures are
not mutually absolutely continuous. So, one may conclude that
$A_{\s}$ does not admit a \qu\ compatible with any Gaussian \rep\ of the
Weyl relations.

We now define the group of linear sympletomorphisms that does admit
a natural \qu\ for a given Gaussian \rep . Let then $C$ be a covariance
operator on $\ss$ and consider the Gaussian \rep\ 
$\Bigl(L^2\bigl(\S'(\R^d),\mu\bigr),\W\Bigr)$ of the Weyl relations,
defined by $C$, where $\m$ is the Gaussian measure of covariance $C$. 
Note that the \rep\ $\W(f,g)$ can be continuously extended to
all $(f,g)\in\H_C\oplus\H_{C^{-1}}$, where $\H_C\oplus\H_{C^{-1}}$ is the real
Hilbert space one obtains by completion of $\S(\R^d)\oplus\S(\R^d)$ with respect to the
inner product 
%%%%%%%%%%%%%%%%%
\be
\label{n4}
\bigl\langle\!\bigl\langle (f,g),(f',g')\bigr\rangle\!\bigr\rangle_C:=
\bigl\langle f,(2C)f\bigr\rangle+\bigl\langle g,(2C)^{-1}g\bigr\rangle\,,
\ee
%%%%%%%%%%%%%%%%%%
as follows from  considerations in Section \ref{appendixB}.

Consider the group of (linear) orthogonal sympletomorphisms, i.e.~the 
group $G_C$ of linear operators $A$ on 
$\H_C\oplus\H_{C^{-1}}$ such that
%%%%%%%%%%%%%%%%%%%%%%%%%
\be
\label{iusl2}
\bigl\langle\!\bigl\langle A(f,g),A(f',g')\bigr\rangle\!\bigr\rangle_C=
\bigl\langle\!\bigl\langle (f,g),(f',g')\bigr\rangle\!\bigr\rangle_C
\ee
%%%%%%%%%%%%%
and
%%%%%%%%%%%%%%%%%%%%%%%
\be
\label{simp}
\bigl\langle A(f,g),\J A(f',g')\bigr\rangle=
\bigl\langle (f,g),\J (f',g')\bigr\rangle\,,
\ee
%%%%%%%%%%%%%%%%
where
%%%%%%%%%%%%%%%%%%%%%%%%%%
\be
\label{bilsimp}
\bigl\langle (f,g),\J (f',g')\bigr\rangle:=
\langle f,g'\rangle - \langle f',g\rangle \,.
\ee
%%%%%%%%%%%%%%%%%%
In what follows we will consider on $G_C$ the topology induced from the strong
topology associated to the $\langle\!\langle\, ,\rangle\!\rangle_C$ norm.
%%%%%%%%%%%%%%%%%%%%%%%%%%%%%%%%%%%%%%%%%%%%%%%%%
\begin{theo}
\label{teoII9.1}
Let $C$ be a covariance operator on $\ss$, $\mu$ the corresponding
Gaussian measure on $\sdd$ and $\W$ the associated Gaussian \rep\
of the Weyl relations. On $L^2\bigl(\S'(\R^d),\mu\bigr)$ there is a unique
strongly continuous unitary \rep\ $U$ of the group $G_C$ such that
%%%%%%%%%%
\be
\label{II9.10}
U(A){\cal W}(f,g)U(A)^{-1}={\cal W}\bigl(A(f,g)\bigr),\ \ \forall A\in G_C,\ \,
\forall f,g\in\S(\R^d)
\ee
%%%%%%%%%%%%
and
%%%%%%%%%%%%%
\be
\label{II9.10a}
U(A)1=1,\ \forall A\in G_C\,.
\ee
%%%%%%%%%
\end{theo}
%%%%%%%%%%%%%%%%%%%%%%%%%%%%%%%%%%%%%%
To prove the theorem, let us start by showing uniqueness. Suppose that $U$ and $\tilde U$ are two
such \reps\ of $G_C$. Then, for all $A\in G_C$, $\tilde U(A)^{-1}U(A)$ 
commutes with $\W(f,g)$, $\forall f,g$, which implies that 
$\tilde U(A)^{-1}U(A)$ is proportional to the identity, given the 
irreducibility of $\W$. 
This still does not prove that $\tilde U$ coincide exactly with $U$, but
since by (\ref{II9.10a}) $\tilde U(A)^{-1}U(A)1=1$, 
we conclude that $\tilde U(A)=U(A)$, $\forall A$. Let us show that a \rep\
exists. For every $A\in G_C$, $\W_A$ defined by  (\ref{iusl1}) is
irreducible and continuous with respect to $\langle\!\langle\, ,\rangle\!\rangle_C$,
since $A$ is orthogonal. The crucial fact is that since the expectation values
$\langle 1,\W(f,g)1\rangle$ depend only on the inner product
$\langle\!\langle\, ,\rangle\!\rangle_C$ (see Section \ref{appendixB}), 
%%%i.e., 
%%%$\langle 1,\W(f,g)1\rangle=\exp\Bigl(-\bigl\langle\!\bigl\langle (f,g),(f,g)
%%%\bigr\rangle\!\bigr\rangle_C/4\Bigr)$ \cite{ref}, 
one gets that
$\langle 1,\W_A1\rangle=\langle 1,\W1\rangle$, $\forall A\in G_C$. This
in turn implies that $\W$ and $\W_A$ are equivalent \reps , $\forall A\in G_C$.
One can show  that the unitary operator $U(A)$ on
$L^2\bigl(\S'(\R^d),\mu\bigr)$ defined by
%%%%%%%%%%%%
\be
\label{iusl4}
\sum_k\lambda_k\W(f,0)1\stackrel{U(A)}{\longmapsto} 
\sum_k\lambda_k\W\bigl(A(f_k,0)\bigr)1,\ \ \lambda_k\in\C
\ee
%%%%%%%%%%%%%%
satisfies both (\ref{II9.10}) and (\ref{II9.10a}). From (\ref{iusl4})
follows immediately that the operators $U(A)$ are a \rep\ of $G_C$.
The proof of the continuity of this \rep\ also presents no significative 
difficulty (see e.g.~\cite{BSZ}).~$\Box$

%%%%%%%%%%%%%%%%%%%%%%%%%%%%%  Section  %%%%%%%%%%%%%%%%%%%%%%%%%%%%%%%%%
\section{Quantization of dynamics: free field measure}
\label{section3}
%%%%%%%%%%%%%%%%%%%%%%%%%%%%%%%%%%%%%%%%%%%%%%%%%%%%%%%%%%%%%%%%%%%%%%%%%

In this section we present the \qu\ of the free real scalar field of mass $m$
in $d+1$ dimensions, following \cite{BSZ}, \cite{RS2} and \cite{GJ}. Recall
that the phase space $\H_m^+\oplus\H_m^-$ and the space of test functions
$\H_m^-\oplus\H_m^+$ for the classical field of mass $m$ are naturally
equiped with a real Hilbert space structure, and that the classical
evolution acts by orthogonal \trfs . Using the inner product on $\H_m^-$ one
can define a Gaussian \rep\ of the Weyl relations for which the quantization 
of the dynamics is guaranteed by theorem \ref{teoII9.1}.

As we saw in section \ref{section1}, proposition \ref{propII1.2.1}, the
operator 
%%%%%%%%%%%%%%%%
\be
\label{II8.1}
C_m:={1\over 2}(m^2-\Delta )^{-1/2}
\ee
%%%%%%%%%%%%%%%
 associated to 
the classical Hamiltonean,
possesses all the properties of a covariance operator on $\ss$.
Let $\mu^m$ be the Gaussian measure on $\sdd$ of covariance $C_m$.
The measure $\m^m $, or equivalently the covariance $C_m$, defines a
(cyclic) Gaussian \rep\ $\Bigl(L^2\bigl(\S'(\R^d),\m^m\bigr),\W_m\Bigr)$
of the Weyl relations such that
%%%%%%%%%%%%%%%%%%%%%%%%%
\be
\label{II8.11}
\!\!\!\!\langle 1,\W_m(f,g)1\rangle \!=\!\exp \Bigl(
-{\textstyle {1 \over 4}}
(\langle f,(m^2-\Delta )^{-1/2}f\rangle+
\langle g,(m^2-\Delta )^{1/2}g\rangle)\Bigr).
\ee
%%%%%%%%%%%%%%%%%%%%%%
Explicitly, the \rep\ is defined by the operators $\U_m(f):=\W_m(f,0)$, 
$\V_m(g):=\W_m(0,g)$:
%%%%%%%%%%%%%%%%%%%%%%%%%%%%%%%
\be
\label{II8.4}
\bigl({\cal U}_m(f)\psi \bigr)(\phi ) =  e^{-i\phi (f)}\psi (\phi )
\ee
%%%%%%%%%%%%%%%%%%%%%%
and
%%%%%%%%%%%%%%%%%%%%%%%%%%%
\be
\label{II8.4a}
\bigl({\cal V}_m(g)\psi \bigr)(\phi )=e^{-\langle g,C^{-1}_mg\rangle /4}\,
e^{\phi(C^{-1}_m g)/2}\,\psi(\phi-g)\, ,
\ee
%%%%%%%%%%%% 
where $\psi\in L^2\bigl(\S'(\R^d),\m^m\bigr)$.

The \qu\ of the coordinate functions
%%%%%%%%%%%%%%
\ba
\label{n5}
\mbox{\boldmath $\varphi$}_f(\vf,\pi)& := & \langle f,\vf\rangle,\ \ f\in\ss \\
\label{n6}
\mbox{\boldmath $\pi $}_g(\vf,\pi)& := & \langle g,\pi\rangle,\ \ g\in\ss 
\ea
%%%%%%%%%%%%%
that generate the kinematical algebra is given by the generators of
$\U_m$ and $\V_m$ as follows. There is a dense subspace ${\cal D}\subset 
L^2\bigl(\S'(\R^d),\m^m\bigr)$ and operators 
$\Q_m(\mbox{\boldmath $\varphi $}_f)$ and $\Q_m(\mbox{\boldmath $\pi $}_g)$ 
(essentially) self-adjoint on $\cal D$ such that \cite[X.7]{RS2}
%%%%%%%%%%%%%%%
\ba
\label{II8.12}
\U_m(sf)&=&\exp \bigl(-is \Q_m(\mbox{\boldmath $\varphi $}_f)\bigr)\,,\\
\label{II8.13}
\V_m(sg)&=&\exp \bigl(-is \Q_m(\mbox{\boldmath $\pi $}_g)\bigr),\ 
\forall s\in\R.
\ea
%%%%%%%%%%%%%%%%%%%%%%
On $\cal D$, the canonical commutation relations
%%%%%%%%%%%%%%%%%%%
\be
\label{II8.14}
[\Q_m(\mbox{\boldmath $\varphi $}_f),\Q_m(\mbox{\boldmath $\pi $}_g)]
=i\langle f,g\rangle{\bf 1}
\ee
%%%%%%%%%%%%%%%%%%%%%%%%%%%%
are satisfied.

As we saw, the \rep\ can be continuously extended to all $(f,g)\in
\H_{C_m}\oplus\H_{C^{-1}_m}$. With $C_m$ given by (\ref{II8.1}), this
Hilbert space is precisely the test function space $\H^-_m\oplus\H^+_m$.
We will stick to this latter notation.
%%%%
%, simplifying also the notation
%for the corresponding inner product. So, the inner product on
%$\H_m^-\oplus\H_m^+$ $(\equiv \H_{C_m}\oplus\H_{C^{-1}_m})$ will be denoted
%from now on by $\langle\!\langle\,,\rangle\!\rangle_m$. With this new
%notation the expectation value (\ref{II8.11}) reads
%%%%%%%%%%%%%%%%%%%%
%\be
%\label{bu}
%\langle 1,\W_m(f,g)1\rangle=\exp\bigl({\textstyle -{1 \over 4}}
%\langle\!\langle(f,g),(f,g)\rangle\!\rangle_m\bigr).
%\ee
%%%%%%%%%%%%%%%%%%%%%%%%%%

The fundamental aspect of the \rep\ $\W_m$ is that of allowing a \qu\
of the dynamics for the field of mass $m$. As we saw in section \ref{section1},
the kinematical observables $F_{c,(f,g)}$, or equivalently the test
functions $(f,g)$, evolve under the action of a group $S_m(t)$ 
(\ref{livre com massa 6}) of orthogonal linear sympletomorphisms on 
$\H^-_m\oplus\H^+_m$. We thus have the following fundamental result, as a
corollary of theorem \ref{teoII9.1}.
%%%%%%%%%%%%%%%%%%%%%%%%%%%%%%%%%%%%%%%%%%%%%%%%%%
\begin{cor}
\label{corII9.1}
On $L^2\bigl(\S'(\R ^d),\m^m\bigr)$ is defined a continuous unitary \rep\
$T_m(t)$ of $\R $ such that
%%%%%%%%%%%%%
\be
\label{II9.38}
T_m(t)\W_m(f,g)T_m(t)^{-1}=\W_m\bigl(S_m(t)(f,g)\bigr),\ \forall t 
\in \R,\ \forall (f,g)
\ee
%%%%%%%%%%%
and 
%%%%%%%%%%%%
\be
\label{II9.39}
T_m(t)1=1,\ \forall t\,.
\ee
%%%%%%%%%%%%%%
\end{cor}
%%%%%%%%%%%%%%%%%%%%%%%%%%%%%%%%%%%%%%%%%%%%%%%%%%%%%%
This result gives a quantization of the dynamics, or 
a quantization $\Q_m(H)$ of the Hamiltonean defined by
%%%%%%%%%%%%%%%%%%%%%%%
\be
\label{quantH}
T_m(t)=:\exp\bigl(it\Q_m(H)\bigr)\,.
\ee
%%%%%%%%%%%%%%%%%%%%%%%%%
It can be shown \cite{BSZ,RS2,GJ} that the quantum Hamiltonean has 
non-negative spectrum and that the zero eigenvalue is non-degenerate,
establishing the interpretation of the cyclic vector 1 as the vacuum.
It can also be shown that the \qu\ of the free field of mass $m$ above
is unique \cite{BSZ}, i.e., given a cyclic \rep\ $\W$ of the Weyl relations
such that there exists an unitary one-parameter group 
$T(t)=:\exp\bigl(it\Q(H)\bigr)$ with non-negative generator $\Q(H)$ 
and such that (\ref{II9.38}) and (\ref{II9.39}) are satisfied, one can find an 
unitary operator $U$ relating both $\W$ to $\W_m$ and $T$ to $T_m$:
%%%%%%%%%%%%%%%%%%%%%%%%%%
\ba
\label{UW}
U\W(f,g)U^{-1}&=&\W_m(f,g)\, ,\ \ \forall f,g\\
UT(t)U^{-1}&=&T_m(t)\, ,\ \ \forall t\, .
\ea
%%%%%%%%%%%%%%%%%%%%%%%%%%%%%%%%%  

%%%%%%%%%%%%%%%%%%%%%%%%%%%%%  Section  %%%%%%%%%%%%%%%%%%%%%%%%%%%%%%%%%
\section{Invariance and ergodicity of the action of the Euclidean group}
\label{section4}
%%%%%%%%%%%%%%%%%%%%%%%%%%%%%%%%%%%%%%%%%%%%%%%%%%%%%%%%%%%%%%%%%%%%%%%%%

In this section we show that the measure $\m^m$ for the free field is
invariant and ergodic with respect to the natural action of the Euclidean group on 
$\R^d$. The invariance gives us an unitary \rep\ of this symmetry group on
the quantum Hilbert space, and ergodicity implies that the vacuum is the
only invariant state (see e.g.~\cite{Ya}). The invariance and ergodicity 
of the measure
$\m^m$, $\forall m$, also implies that two measures $\m^m$ and $\m^{m'}$,
$m\neq m'$, are supported on disjoint sets (see e.g.~\cite{Ya}), which in 
turn leads 
immediately to the nonequivalence of the corresponding \reps\ of the
Weyl relations.

The Euclidean group $\E$ on $\R^d$ acts on $\S(\R^d)$ by
%%%%%%%%%%%%%%%%
\be
\label{II11.1}
f\mapsto f_{\gamma}:f_{\gamma}(x)=f(\gamma^{-1}x)\,,
\ee
%%%%%%%%%%%%%%%%%%
where $\gamma\in\E$ and $\gamma x$ denotes the natural action of $\E$
on $\R^d$. The action on $\ss$ induces an action $\vf$ on $\sdd$:
%%%%%%%%%%%%%%%%%%
\be
\label{II11.2}
\bigl(\varphi _{\gamma }\phi \bigr)(f)=\phi (f_{\gamma })\,.
\ee
%%%%%%%%%%%%%
The invariance of the measure $\m^m$ follows immediately from the invariance 
of the covariance $C_m={1\over 2}(m^2-\Delta)^{-1/2}$.
%%%%%%%%%%%%%%%%%%%%%%%%%%%%%%%%%%%%%%%%%%%%%%%%%%%%%%
\begin{prop}
\label{propesr1}
The measure $\m^m$ is $\E$-invariant, for any $m$.
\end{prop}
%%%%%%%%%%%%%%%%%%%%%%%%%%%%%%%%%%%%%%%%%%%%%%%%%%%%%%%%
One thus have an unitary action $U$ of $\E$ on $L^2\bigl(\sdd,\m^m\bigr)$:
%%%%%%%%%%%%%%%%
\be
\label{n7}
\bigl(U(\gamma)\psi\bigr)(\phi)=\psi\bigl(\vf^{-1}_{\gamma}\phi\bigr),
\ \ \psi\in L^2\bigl(\sdd,\m^m\bigr)\,.
\ee
%%%%%%%%%%%%%%%%%
Let us consider the subgroup (isomorphic to $\R$) of $\E$ of all
translations in a fixed direction (for instance parallel to the $x_1$ axis):
%%%%%%%%%%%%%%%%%
\be
\label{II11.3}
(x_1,x_2,\ldots,x_d)\longmapsto (x_1+y,x_2,\ldots,x_d),\ y\in \R \,.
\ee
%%%%%%%%%%%%%%%%%%
%%%%%%%%%%%%%%%%%%%%%%%%%%%%%%%%%%%%%%%%%%%%%%
\begin{deff}
\label{defesr1}
Given a probability space $M$ with measure $\m$, a measure preserving
action of $\R$ is said to be mixing if the corresponding unitary action $U$ on 
$L^2(M,\mu )$ satisfies
%%%%%%%%%%%%%
\be
\label{II11.4}
\lim_{y\to \infty }\langle\psi ',U(y)\,\psi \rangle=\langle\psi ',1\rangle
\langle1,\psi \rangle,\ \forall \psi ',\psi \in L^2(M,\mu )\,.
\ee
%%%%%%%%%%%%%
\end{deff}
%%%%%%%%%%%%%%%%%%%%%%%%%%%%%%%%%%%%%%%%%%
It follows from (\ref{II11.4}) that any invariant element of
$L^2(M,\mu )$ is constant a.e., and therefore mixing implies ergodicity
\cite{RS1,Si}.
%%%%%%%%%%%%%%%%%%%%%%%%%%
\begin{prop}
\label{propesr2}
The action (\ref{n7}) of the subgroup (\ref{II11.3}) is mixing.
\end{prop}
%%%%%%%%%%%%%%%%%%%%%%%%%
In fact, by linearity and continuity, it is sufficient to verify (\ref{II11.4})
for the functions of the form $e^{-i\phi (f)}$, whose linear span is dense.
For those functions:
%%%%%%%%%%%%%%%%%%%%
\ba
\label{II11.5}
& \bigl\langle &\!\!\!\!\!
e^{-i\phi (f')}, U(y)\,e^{-i\phi (f)}
\bigr\rangle\,= 
\nonumber \\
& = &\!\!\! 
\int e^{-i\phi (f_{-y}-f')} d\mu ^m \nonumber\\
\label{II11.6}
& = &\!\!\! 
\exp \Bigl( -\textstyle {1 \over 2}\bigl\langle (f_{-y}-f'),
C_m(f_{-y}-f')
\bigr\rangle\Bigr) \nonumber\\
\label{II11.7}
& = &\!\!\! 
\bigl\langle e^{-i\phi (f')},1\bigr\rangle \bigl\langle 1,
e^{-i\phi (f)}\bigr\rangle\, \cdot \\
& \ &\!\!\! 
\cdot\, \exp \Bigl( \textstyle {1 \over 2}\int f'(x_1,\ldots,x_d)
(m^2-\Delta )^{-1/2}f(x_1+y,\ldots,x_d)d^dx  \Bigr)\nonumber . 
\ea
%%%%%%%%%%%%%%%%%%%
By Fourier transform, the integral in exponent on (\ref{II11.7}) can be
written as
%%%%%%%%%%%%%%%%%
\be
\label{II11.8}
\int e^{-ik_1y}\left(\int{\tilde{f'}(k)\tilde{f}(k)\over (m^2+k^2)^{1/2}}
\, dk_2\cdots dk_d\right)dk_1\ .
\ee
%%%%%%%%%%%%%%%%%%
The integral (\ref{II11.8}), seen as a function of $y$, is the Fourier
transform of a function in $\S(\R)$. By the Riemann-Lebesgue lemma,
(\ref{II11.8}) goes to zero in the limit $y\to \infty$. Going back to
(\ref{II11.7}), we conclude that the action is mixing.~$\Box$

The subgroup (\ref{II11.3}), and consequently any subgroup that contains it,
acts therefore ergodically. Note also that the $\E$-invariance of the 
measure implies that the subgroup of translations in any fixed direction acts
ergodically. We conclude that: {\it i\/}) the vacuum is the only state
invariant under the action of translations of the type (\ref{II11.3});
{\it ii\/}) two measures $\m^m$ and $\m^{m'}$, $m\neq m'$, are supported
on mutually disjoint sets, implying the non-unitary equivalence of the
corresponding free-field \reps.

%%%%%%%%%%%%%%%%%%%%%%%%%%%%%%%%%
\section{Covariant formulation}
\label{section5}
%%%%%%%%%%%%%%%%%%%%%%%%%

We will   now  consider the relativistic invariance properties of the free field quantization,
showing explicitely how  a covariant formulation can be obtained from the above canonical quantization.
Although presented here in heuristic form, one can give a precise meaning to 
the results in this section (see \cite[6.3]{BSZ} for a rigourous approach).
We start by considering the evolution of the quantum operators 
$\Q_m(\mbox{\boldmath $\varphi $}_f)$ and $\Q_m(\mbox{\boldmath $\pi $}_g)$
(see eqs. (\ref{II8.12}), (\ref{II8.13})), which is simply dictated by the
classical evolution of test functions. One can then fully reconstruct a
relativistic quantum field obeying an appropriate quantum version of
the classical Klein-Gordon equation.

Let us consider the time-dependent Weyl operators: 
%%%%%%%%%%%%%%%%
\be
\label{II10.1}
\W_m^t:=T_m(t)\W_m (f,g)T_m(t)^{-1}=\W_m \bigl(S_m(t)(f,g)\bigr)\,.
\ee
%%%%%%%%%%%%%%%
To simplify the notation we introduce time-dependent field operators
$\hat \vf_t(f)$ and $\hat \pi_t(g)$ such that 
$\hat \vf_0(f)=\Q_m(\mbox{\boldmath $\varphi $}_f)$ and
$\hat \pi_0(g)=\Q_m(\mbox{\boldmath $\pi $}_g)$.
These operators describe the evolution of the corresponding time-zero
operators and are defined by the following condition:
%%%%%%%%%%%%%%%
\be
\label{r1}
\W_m^t(sf,rg)=\exp[-is\hat \vf_t(f)-i\hat \pi_t(g)],\ \forall s,r\in\R, \
\forall f,g\in\S(\R^d).
\ee
%%%%%%%%%%%%%%
We can rewrite this expression in the more convenient form
%%%%%%%%%%%%%%%
\be
\label{r2}
\W_m^t(sf,rg)=\exp\left [
-i(\hat \vf_t\ \hat \pi_t)
\left (\begin{array}{c}
sf\\ rg
\end{array}
\right )
\right ]\, ,
\ee
%%%%%%%%%%%%%%
where 
$(\hat \vf_t\ \hat \pi_t)
\left (\begin{array}{c}
f\\ g
\end{array}
\right )$
stands for $\hat \vf_t(f)+\hat \pi_t(g)$. 
Equation (\ref{II10.1}) then translates to
%%%%%%%%%%%%%%%
\be
\label{r3}
\W_m^t(sf,rg)=\exp\left [
-i(\hat \vf_0\ \hat \pi_0)S_m(t)
\left (\begin{array}{c}
sf\\ rg
\end{array}
\right )
\right ]\, .
\ee
%%%%%%%%%%%%%%
From (\ref{r2}) and  (\ref{r3}) follows that:
%%%%%%%%%%%%%%%%
\ba
\label{r4}
\hat \vf_t(f) & = & (\hat \vf_0\ \hat \pi_0)S_m(t)
\left (\begin{array}{c}
f\\ 0
\end{array}
\right )\,, \\
\label{r5}
\hat \pi_t(g) & = & (\hat \vf_0\ \hat \pi_0)S_m(t)
\left (\begin{array}{c}
0\\ g
\end{array}
\right )\,.
\ea
%%%%%%%%%%%%%%%%%%%%%%
To arrive at the relativistic formulation we will obtain first the second
order differential equation for $\hat \vf_t(f)$.
Taking into 
account that (see section \ref{section1})
%%%%%%%%%%%%%%%%%
\be
\label{II10.4}
{dS_m(t)\over dt}=S_m(t)\left( \begin{array}{cc} 0 & -(m^2-\Delta ) \\
1 & 0 \end{array}\right)\, ,
\ee
%%%%%%%%%%%%%%%
one gets, after using again (\ref{r4}) and (\ref{r5}):
%%%%%%%%%%%%%%%%%%%%%%%
\ba
\label{r6}
{d \over dt}\,\hat \vf_t(f) & = & \hat \pi_t(f)\,, \\
\label{r7}
{d \over dt}\,\hat \pi_t(g) & = & - \,\hat \vf_t\bigl((m^2-\Delta )g\bigr)\,.
\ea
%%%%%%%%%%%%%%%%%%%
The second order equation now follows:
%%%%%%%%%%%%%%%%%%%%%%%%
\be
\label{II10.7}
{d^2\over dt^2}\,\hat \vf_t(f)+\hat \vf_t\bigl((m^2-\Delta )f\bigr)=0\,.
\ee
%%%%%%%%%%%%%%%%%%%%%%%%%%
We now introduce space-time averages. 
For any  $F(t,x)\in\S(\R^{d+1})$, the integral
%%%%%%%%%%
$$
\int \hat\vf_t\bigl(F(t,\cdot)\bigr)dt
$$
%%%%%%%%%%%
defines a self-adjoint operator \cite[6.3]{BSZ}, which we will denote by
{\boldmath $\Phi$}$(F)$.
The map {\boldmath $\Phi $} from $\S(\R^{d+1})$ to self-adjoint  operators 
on $L^2\bigl(\S'(\R^d),\m^m\bigr)$ is interpreted as the relativistic 
quantum field, and satisfies an equation analogous to the classical
Klein-Gordon equation:
%%%%%%%%%%%%%%%%%%%%%
\be
\label{II10.14}
\mbox{\boldmath $\Phi $}\bigl((\Box +m^2)F\bigr)=0\,.
\ee
%%%%%%%%%%%%%%%%%%
One can show that there exists a unitary \rep\  $\Gamma $ of the 
Poincar\'e group such that for any  Poincar\'e transformation $\Lambda$ one has
%%%%%%%%%%%%%%%%
\be
\label{II10.18}
\Gamma (\Lambda )1=1
\ee
%%%%%%%%%%%%%%%%%%%%%%%
and
%%%%%%%%%%%%%%%%%%%%%
\be
\label{II10.17}
\Gamma (\Lambda )\mbox{\boldmath $\Phi $}(F)\Gamma (\Lambda )^{-1}=
\mbox{\boldmath $\Phi $}(F_{\Lambda })\,,
\ee
%%%%%%%%%%%%%%
where $F_{\Lambda }(t,x)=F\bigl(\Lambda ^{-1}(t,x)\bigr)$.

\section{Local properties of the support of the free field measure}
\label{s2}

We will now present a characterization of the support of the free field measure $\m^m$ defined by the covariance 
  (\ref{II8.1}), following \cite{RR,CL,Ri} and \cite{ZeS}.
The result in question is a consequence of the so-called Minlos' theorem,
which for  the case of Gaussian measures in $\S'(\R^d)$
can be stated as follows (see \cite{RR,Ri,Ya}): 

%[We will use the term {\em covariance} to denote also the scalar product $(\, ,)$ that defines the measure.]

\bt[Minlos]
\label{minlos}
Let   $(\,,)$ be a continuous inner product  on $\S(\R^d)$
and $\H$
the corresponding completion of $\S(\R^d)$. Let $H$ be an injective Hilbert-Schmidt operator
on the Hilbert space $\H$, such that $\S(\R^d)\subset H\H$ and $H^{-1}:\S(\R^d)\to\H$ is a continuous map.
Let $(\,,)_1$ be the inner product on
 $\S(\R^d)$ defined by $(f,g)_1=(H^{-1}f,H^{-1}g)$.Then, the
Gaussian measure on the dual space $\S'(\R^d)$ with covariance $(\,,)$ is supported on
the  subspace of 
$\S'(\R^d)$ of those functionals which are continuous with respect to the topology defined by  
$(\,,)_1$.
\et

%%%%%%%%%%%%%%%%%%%%%%%%%%%%%%%%%%%%%%%%%%%%%%%%%%%%%%%%%%%%%%%%%%%%%%%%%%%%
%%%%%%%%%%%%%%%%%%%%%%%%%%%%%%%%%%%%%%%%%%%%%%%%%%%%%%%%%%%%%%%%%%%%%%%%%%%%
%%%%%%%%%%%%%%%%%%%%%%%%%%%%%%%%%%%%%%%%%%%%%%%%%%%%%%%%%%%%%%%%%%%%%%%%%%%
%%%%%%%%%%%%%%%%%%%%%%%%%%%%%%%%%%%%%%%%%%%%%%%%%%%%%%%%%%%%%%%%%%%%%%%%%%%%
Let us then apply Theorem \ref{minlos} to the Gaussian measure
defined by the covariance operator (\ref{II8.1}), so that we can obtain sets of measure one.
We will show that the set of distributions  supporting the measure are such that 
the action of the operator $(1+x^2)^{-\a}(m^2-\Delta)^{-\beta}$ produces 
$L^2(\R^d)$ elements, for $\a>d/4$ e $\beta>(d-1)/4$. To prove this, let us
consider the operators
%%%%%%%%%%%
\be
\label{clivre1a}
\widetilde H:=(m^2-\Delta)^{-\a}(1+x^2)^{-\a}
\ee
%%%%%%%%%%%
on $L^2(\R^d)$, where $(1+x^2)$ is a   multiplication operator and
$\a>d/4$. Since $(1+x^2)^{-\a}$ is square integrable and the same is true for the
 Fourier transform $(m^2+p^2)^{-\a}$ of $(m^2-\Delta)^{-\a}$,
the operators $\widetilde H$ are of the Hilbert-Schmidt type
$\forall\a >d/4$. 
Like in Section \ref{section1}, let  $\H^-_m$ be  the completion of 
$\S(\R^d)$ with respect to the inner product (\ref{app3}) associated with the operator (\ref{II8.1}).
Taking advantage of the unitary transformation 
%%%%%%%%%%%%%
\be
\label{eqclivre2}
(m^2-\Delta)^{1/4}:L^2(\R^d)\to\H^-_m,
\ee
%%%%%%%%%%%%%%
one can define  an Hilbert-Schmidt $H$ operator on $\H^-_m$: 
%%%%%%%%%%%%%
\be
\label{clivre1b}
H=(m^2-\Delta)^{1/4}\widetilde H(m^2-\Delta)^{-1/4}\,.
\ee
%%%%%%%%%%%%
Let us finally  introduce the following inner product in $\S(\R^d)$: 
%%%%%%%%%%%%%%
\ba
\label{clivre1c}
(f,g)_1:& = &\int d^d x  \left((m^2-\Delta)^{-1/4}H^{-1}f\right)
\left((m^2-\Delta)^{-1/4}H^{-1}g\right) \nonumber \\
& = & \int d^d x  \left((1+x^2)^{\a}(m^2-\Delta)^{\beta}f\right)
\left((1+x^2)^{\a}(m^2-\Delta)^{\beta}g\right)\,,
\ea
%%%%%%%%%%%%%%
where $\beta=\a-1/4>(d-1)/4$. 
%Let $\widehat\H^-_m\subset\H^-_m$ be the completion of
% $\S(\R^d)$ with respect to the scalar product $(\,,)_1$. 
 By Theorem \ref{minlos}, the subspace 
of those  functionals which are continuous with respect to $(\,,)_1$ is a set of measure one. 
The support of the measure can therefore be written as follows:
\be
\label{minlosup}
(m^2-\Delta)^{\b}(1+x^2)^{\a} L^2(\R^d),
\ee
%%%%%%%
in the  sense that the distributions which support the measure
are such that the application of the operator 
$(1+x^2)^{-\a}(m^2-\Delta)^{-\b}$ produces elements of $L^2(\R^d)$, for $\a>d/4$ and $\beta>(d-1)/4$. 

So, 
one can say that the
Fourier transform of $(m^2+p^2)^{-\b}\widetilde\phi(p)$ is locally $L^2$,
for almost every distribution $\phi(x)$, where $\widetilde\phi(p)$
denotes the  Fourier transform.  Further application of the operator $(1+x^2)^{-\a}$ regularizes the behaviour at infinity of typical
distributions, producing truly $L^2$ elements.

Note that although the value of mass $m$ appears explicitly in the characterization of the support
given by 
(\ref{minlosup}), the space that one obtains for support of the measure as a consequence of Minlos'
theorem 
is actually the same for all values of the mass. 
This is a consequence of the fact that the topology defined by the scalar product
(\ref{clivre1c}) is independent of the (nonzero) value of the  mass.
So, the above description of the support of the measure is not sensitive to the value of the mass
of the free field.
%Ultimately, this can be traced back to the fact that $(m^2-\Delta)^{-1/2}({m'}^2-\Delta)^{1/2}$ is a bounded
%operator, of bounded inverse, $\forall m,m'$.

Nevertheless, as mentioned in Section \ref{section4}, the measures associated with two 
distinct values of the mass are in fact singular with respect to each other.
Therefore, disjoint supports  can be found, for distinct masses.
To disclose these crucial differences in the support requires a different type
of analysis, namely one that takes into account  the large scale behaviour of typical distributions.
We address this question in the next section.
%which we address on the next section section, following \cite{MTV} (with the necessary adaptations
%to the canonical setting

\section{Long range behaviour: distinction of the supports for different values of the mass}
\label{clivre2}
%%%%%%%%%%%%%%%%%%%%%%%%%%%%%%%%%%%%%%%%%%%%%%%%%%%%%%%%%%%%%%%%%%%%%%%%%%%%%%%
%%%%%%%%%%%%%%%%%%%%%%%%%%%%%%%%%%%%%%%%%%%%%%%%%%%%%%%%%%%%%%%%%%%%%%%%%%%%%

%two measures $\m^m$ and $\m^{m'}$, $m\neq m'$, are supported
%on mutually disjoint sets

It is well known that the free field measures  $\m^m$ are singular with respect to each other for different values of the mass.
In order to distinguish the supports of the measures corresponding to different masses we will now analyze 
the long range behaviour of typical distributions, which is sensitive
to the value of the mass. We adopt here the same method as in \cite{MTV}, with the difference that we are considering now the measure defined by the covariance (\ref{II8.1}), featuring in the canonical quantization
approach, instead of  the corresponding Euclidean path integral measure considered in \cite{MTV}.

In the inverse covariance $C_m^{-1}=2(m^2-\Delta)^{1/2}$ one can identify a
diagonal term, which favours a white noise type of behaviour, and  a
nondiagonal term which imposes  correlations between different regions in space, which however decay with distance. So, one can expect that the typical quantum field will present strong correlations
for small distances and will approach white noise behaviour at large distances. The scale distance
is clearly marked by the value of the mass, i.e. $m^{-1}$ can be interpreted as 
a correlation lenght. The expected behaviour at scales much larger than $m^{-1}$
is  therefore that of a white noise type of measure with covariance $\sigma=(2m)^{-1}$.

In order to obtain our formal result, let us consider the  measurable functions 
$F_{B_j}:\S'(\R^d)\to \C$ given by 
%%%%%%%%%%%%%%%%%
\be
\label{II.11.25}
F_{B_j}:\ \phi \mapsto F_{B_j}(\phi)\equiv
\phi(f_j) = {1 \over L^d} \int_{B_j} \phi(x) d^dx\,,
\ee
%%%%%%%%%%%%%%%%%%
where $\{B_j\}_{j=1}^{\infty}$ is a family of mutually disjoint hypercubes in 
$\R ^d$, of edge length  $L$, and  
 $f_j$ denotes the characteristic function of the hypercube $B_j$ multiplied by $1/L^d$. 
We will consider the family of hypercubes 
$\{B_j\}_{j=1}^{\infty}$ centered at points
$x^j =(x_1^j,x_2^j\ldots, x^j_d) =({j^2 \over m} , 0, ... , 0)$ and with faces paralel 
to the coordinate planes.  

The push-forward of the free field measure $\m^m$ with respect to the map
%%%%%%%%%%%%%%%%%%%%%%%%%
\ba
\label{II.11.31}
\sdd &\rightarrow& \R^{\Ni}  \nonumber \\
\phi &\mapsto& \{\phi(f_j)\}  
\ea
%%%%%%%%%%%%%%%%%%
is the Gaussian measure $\nu_m$ in $\R^{\Ni}$ with covariance matrix
$\M_m$ given by
%%%%%%%%%%%%%%%%%%%%%%%%%
\be
\label{II.11.32}
(\M_m)_{jl}=\langle f_j,C_mf_l\rangle\, .
\ee
%%%%%%%%%%%%%%%%%%%%%%%%
The matrix  elements can be easily found by  Fourier transform. We get 
%%%%%%%%%%%%%%%%%%%%%%%%%
\be
\label{II.11.33}
(\M_m)_{jl}={1\over 2}\left({2 \over \pi}\right)^d{1 \over L^{2d}}
\int_{\R^d} d^d k\,  e^{i{k_1 \over m}(j^2 - l^2)}
{1 \over (k^2 + m^2)^{1/2}} \prod_{n=1}^d{{\sin}^2(k_nL/2) \over k_n^2}\, .
\ee
%%%%%%%%%%%%%%%%%%%%%%%%
The Fourier transform in (\ref{II.11.33}) is well defined, since 
%%%%%%%%%%%%%%%
$$
{1 \over (k^2 + m^2)^{1/2}} \prod_{n=1}^d{{\sin}^2(k_nL/2) \over k_n^2}
$$
%%%%%%%%%%%%%%
is absolutely integrable in $\R^d$, which also shows that $f_j\in\H^-_m$,
$\forall j$, proving therefore that the map (\ref{II.11.31}) is well defined
 (see Section \ref{appendixA}). As expected, due to the invariance of the measure with 
 respect to spatial translations, the diagonal elements of the matrix
$\M_m$ are all equal. Let us denote them by $\lambda_m^{L}$: 
%%%%%%%%%%%%%%%
\ba
\label{II.11.34}
\lambda_m^{L} &:=& (\M_m)_{jj} =\\
&=&{1\over 2}\left({2 \over \pi}\right)^d{1 \over L^{2d}}
\int_{\R^d} d^d k\, 
{1 \over (k^2 + m^2)^{1/2}} 
\prod_{n=1}^d{{\sin}^2(k_nL/2) \over k_n^2} \nonumber
\ea
%%%%%%%%%%%%%%%%%
and let $\mu_{\lambda^L_m}$ be the Gaussian measure in $\R^{\Ni}$ of diagonal covariance 
matrix   $\lambda^L_m {\bf 1}$ (where ${\bf 1}$ is the diagonal matrix in $\R^{\Ni}$).
%%%%%%%%%%%%%%%%%%%%%%
\begin{lem}
\label{lema1}
The measures $\nu_m$ and $\mu_{\lambda^L_m}$
mutually absolutely continuous, i.e.~have the same zero measure sets.
\end{lem}
%%%%%%%%%%%%%%%%%%%%%%
To prove the lemma we will rely on Theorem I.23 of \cite{S} (see also Theorem 10.1
of \cite{Ya}), which gives necessary and sufficient conditions for two covariances to give
rise to mutually absolutely continuous Gaussian measures. In the present case, since the covariance
of $\mu_{\lambda^L_m}$ is proportional to the identity, it is sufficient to show that
%Usamos o teorema \ref{medidas equiv} com $E=\R^{\Ni}_{\rm c}$ (ver sec\cao\ 
%\ref{mgauss}), $(x,y)=\lambda^L_m\langle x,y\rangle:=\lambda^L_m\sum x_iy_i$ e
%$(x,y)_1=\langle x, \M_m y\rangle $. Dada a rela\cao\ entre 
%$(\, ,\, )$ e o produto interno $\ell^2$, as condi\coes\ ({\it i}\/) e 
%({\it ii}\/) do teorema \ref{medidas equiv} traduzem-se em 
({\it i}\/) $\M_m$ is bounded, positive with bounded inverse in
 $\ell^2$ and ({\it ii}\/) 
$T:=\M_m-\lambda^L_m{\bf 1}$ is Hilbert-Schmidt in $\ell^2$. That $\M_m$ is 
positive and injective follows from the fact that  $\M_m$ is the restriction
of $C_m$ to the  linearly independent sistem $\{f_j\}_{j\in\Ni}$. 
Let us admit for a moment that ({\it ii}\/) is proved. It is then 
clear that $\M_m$ is bounded, since it is the sum of two bounded operators.
Let us suppose that $\M_m$ does not have a bounded inverse. 
Then, by definition, $-\lambda^L_m$ belongs to the spectrum of
$T$. But $T$ is compact and therefore the (nonzero) points of the spectrum are proper 
values, which contradicts the injectivity of  $\M_m$.
Thus, it remains only to show that $T:=\M_m-\lambda^L_m{\bf 1}$ 
is Hilbert-Schmidt. The matrix elements of $T$ are $T_{jj}=0$ and
$T_{jl}=(\M_m)_{jl}$ for $j\not = l$. One can conclude from (\ref{II.11.33})
that the nondiagonal elements  $(\M_m)_{jl}$ are the values at points $j^2-l^2$ of the Fourier transform of a real function $f$.
In fact, taking into account the change of variable  $p_n=k_n/m$ we get
%%%%%%%%%%%%%%%%%%%%%%%%% 
\be
\label{II.11.35}
(\M_m)_{jl}=\int_{\R} dp_1\, e^{ip_1(j^2-l^2)} f(p_1)\ ,
\ee
%%%%%%%%%%%%%%%%%%%%%%%%%%
where
%%%%%%%%%%%%%%%%%%%%%%%%%
\ba
\label{II.11.36}
f(p_1)&:=&
{1\over 2}\left({2 \over \pi}\right)^d{1 \over m^{d+1} L^{2d}}
{{\sin}^2(mLp_1/2) \over p_1^2}\ \cdot\nonumber \\
&\cdot&\int_{\R^{d-1}} d^{d-1}p\, {1 \over (1+p_1^2+\sum_2^d p_n^2)^{1/2}} 
\prod_{2}^d{{\sin}^2(mLp_n/2) \over p_n^2}\,.
\ea
%%%%%%%%%%%%%%%%%%%%%%%%%%%%%%%%%%%%%%%%%%%%%%%%
The above function  $f$ is clearly diferenciable and its derivative $f'$ 
belongs to $L^1(\R)$. We therefore obtain for the  
Fourier transform $\tilde f$:
%%%%%%%%%%%%%%%%%%%
\be
\label{II.11.36b}
\tilde f(p_1)={\widetilde{f'}(p_1) \over ip_1}\ ,
\ee
%%%%%%%%%%%%%%%%%%%%%%%%% 
with $\widetilde{f'}$ being continuous, bounded  and approaching zero at infinity.
One can therefore find $A$, $0<A<\infty$, such that
%%%%%%%%%%%%%%%%%%%%%%%%
\be
\label{II.11.37}
| \tilde f(p_1)|\leq{A \over |p_1|}\ .
\ee
%%%%%%%%%%%%%%%%%%%%%%%%%%%%%%%%%% 
Passing to the matrix elements  $(\M_m)_{jl}$ we get
%%%%%%%%%%%%%%%%%%%%%%%%%%%%%% 
\be
\label{II.11.38}
|(\M_m)_{jl}|^2 \leq {A \over (j^2-l^2)^2}\ , \ \ {\rm for}\ j\not = l\ .
\ee
%%%%%%%%%%%%%%%%%%%%%%%%%%%%%
Using the standard $\ell^2$ basis $e_k=(\delta_{kn})$ we finally obtain 
%%%%%%%%%%%%%%%%%%%%%%%%%%%%%% 
\be
\label{II.11.39}
\sum_l\langle Te_l,Te_l\rangle=\sum_{j,\, l} |T_{jl}|^2 \leq A \sum_{j\not = l}
{1 \over (j^2-l^2)^2} < \infty\ ,
\ee
%%%%%%%%%%%%%%%%%%%%%%%%%%%%
showing that $T$ is Hilbert-Schmidt.~$\Box$
\smallskip
\begin{prop}
\label{resultado}
Let $\mu_{\rho}$ be the Gaussian measure in $\R^{\Ni}$ of diagonal covariance 
$\rho {\bf 1}$, with $\rho>0$. Consider the  measurable subsets of $\R^{\Ni}$
defined by
\be
\label{II.11.12}
Z_{\rho}^{\epsilon}
:=\bigl\{x\ |\ \ \exists N_x\in\N \quad \mbox{\rm such that}\quad 
|x_n|<\sqrt{2(1+\epsilon)\rho\ln n}\,,\ \mbox{\rm for}\ n\geq N_x\bigr\}\,,
\ee
%%%%%%%%%%%%%%%
where $\epsilon\geq 0$. Then the sets   $Z_{\rho}^{\epsilon}$ have $\mu_{\rho}$ measure equal to one for all \mbox{$\epsilon>0$}. 
The set $Z_{\rho}^0$ has $\mu_{\rho}$ measure equal to zero.
\end{prop}
This proposition can be proven as follows.  Consider the measurable
sets $Z_{\rho}^{\epsilon}(N)$, $N\in\N$ defined by
%%%%%%%%%%%%%%%%%%%%%%%%%
\be
\label{II.11.13}
Z_{\rho}^{\epsilon}(N):=\bigl\{x\ |\ \ |x|<\sqrt{2(1+\epsilon)\rho\ln n},
\quad \mbox{\rm para}\quad n\geq N\bigr\}.
\ee
%%%%%%%%%%%%%%%%%%%
Since the measure $\ms$ is in fact a product measure of identical Gaussian measures in $\R$,
it is not difficult to see that the  $\ms$ measure of this sets is
%%%%%%%%%%%%%%%%%%%%%%%%%
\be
\label{II.11.14}
\Lambda_N^{\epsilon}:=\ms\bigl(Z_{\rho}^{\epsilon}(N)\bigr)=
\prod^{\infty}_{n=N}\mbox{Erf}\bigl(
\sqrt{(1+\epsilon)\ln n}\bigr)\,,
\ee
%%%%%%%%%%%%%%%%%%%%%
where Erf$(x) = 1/\sqrt{\pi} \int_{-x}^x e^{-\xi^2} d\xi$ is the error function. 
Given that
$$Z_{\rho}^{\epsilon}(N)\subset Z_{\rho}^{\epsilon}(N+1)$$
and 
$$Z_{\rho}^{\epsilon}=\bigcup_{N\in\Ni}Z_{\rho}^{\epsilon}(N)\ ,$$
it follows from the $\sigma$-additivity of the measure that 
%%%%%%%%%%%%%%%%%%%%%%%%%
\be
\label{II.11.16}
\label{3.13}
\ms \bigl(Z_{\rho}^{\epsilon}\bigr)=\lim_{N\to\infty}\,\Lambda_N^{\epsilon}\,.
\ee
%%%%%%%%%%%%%%%%%%%%%
Since, in particular, the infinite product $\Lambda_2^{\epsilon}$ exists
$(0\leq\Lambda_2^{\epsilon}\leq 1)$ we have that $\lim_{N\to\infty}\,\Lambda_N^
{\epsilon}=1$ or $\Lambda_2^{\epsilon}=0$. The condition
$\Lambda_2^{\epsilon}=0$ is equivalent to $\Lambda_N^{\epsilon}=0$, 
$\forall N>1$, given that $\mbox{Erf}\bigl(\sqrt{(1+\epsilon)\ln n}\bigr)>0$,
$\forall n>1$. So, $\ms(Z_{\rho}^{\epsilon})$ is equal to zero if
$\Lambda_2^{\epsilon}=0$ and equal to one otherwise. Passing to
logaritms, it remains to consider the convergence of the infinite sum 
%%%%%%%%%%%%%%%%%%%%%%%%%
\be
\label{II.11.17}
\sum^{\infty}_{n=2}\ln\bigl(\mbox{Erf}(\sqrt{\ln n^{1+\epsilon}})\bigr)\,.
\ee
%%%%%%%%%%%%%%%%%%%%%%
Replacing the general  term  $\ln\bigl(\mbox{Erf}(\sqrt{\ln n^{1+\epsilon}})
\bigr)$ in (\ref{II.11.17}) by its  asymptotic limit, namely
$-1/\bigl(n^{1+\epsilon}\sqrt{(1+\epsilon)\ln n}\bigr)$, we conclude that the sum
 (\ref{II.11.17})  diverges for $\epsilon=0$ and is convergent
 $\forall\epsilon>0$.~$\Box$

\smallskip
Finally, by combining lemma \ref{lema1} with proposition \ref{resultado}, we immediately
obtain the following result concerning the support of the free field measures $\m^m$:
%%%%%%%%%%%%%%%%%%
\bt
\label{Yme}
The sets
%%%%%%%%
\ba
\label{II.11.40}
Y^{\epsilon}_m&:=&\{ \phi \in \sdd\ |\ \ \exists N_{\phi}\in\N 
\ \ \hbox{\rm such that}\nonumber \\  
&|\phi(f_n)|& < 
\sqrt{2 (1+\epsilon)\lambda_m^{L} \ln n},\ \ 
\mbox{\rm for}\,\ n \geq N_{\phi} \} 
\ea
%%%%%%%%%
have  $\m^m$ measure one for any $\epsilon >0$ and measure zero
for $\epsilon =0$. For 
$m'<m$ there exists  $\epsilon (m')>0$ such that
$\mu^{m'}\bigl(Y^{\epsilon(m')}_m\bigr)=0$.
\et
%%%%%%%%%%%%%%%%%%%%%%%
The second statement of the theorem, which in particular allows the construction
of disjoint supports for  $m\not =m'$, follows from the strict monotony of
$\lambda^L_m$ as a function of $m$.~$\Box$
%; em particular $m'<m$ implica $\lambda^L_{m'}>\lambda^L_m$

\section{Conclusions}
Measures in infinite dimensional spaces, both linear and nonlinear, are of major importance 
in quantum theory. They appear already in standard Quantum Mechanics, in the path integral formulation,
but are even more relevant in the quantum theory of  fields, both from the 
Euclidean path integral and from the canonical quantization perspectives (see e.g.~\cite{GJ}), 
including the quantization of gravity (see  \cite{AL,T}).
These measures are typically defined
on extensions of the spaces associated with the  classical field theories in question.
 In the case of scalar fields one finds measures in the  space
$\S'(\R^d)$ of tempered distributions,  dual to the  
  Schwartz space. 
  %In quantum gravity, on the other hand, one finds measures in the so-called space of generalized connections
  % \cite{AL,T}, an  extension of the space of connections on a manifold.
 In all known cases the distributional extensions are crucial, given that, for measures of
 interest, the classical non-distributional configurations  turn out to be  (subsets of) sets  of  measure zero.

The  construction of measures corresponding to field theories with interactions  
turned out to be a problem of major   complexity, in
particular for spacetime dimensions above $1+1$ (see nevertheless
\cite{GJ,Ri,S,RS2,RS3}). The quantum theory of the free scalar field (in arbitrary   $d+1$ dimensions),
however, is well established  and has been extensively studied, under several viewpoints.  
The simplification resides, of course, in the  linearity of the dynamics: 
the free field Hamiltonean is quadratic,  allowing a quantization of the model by means
of  well defined  Gaussian measures.
In the canonical  formalism, the free field in $d+1$
dimensions is quantized using a Gaussian measure in $\S'(\R^d)$ which is effectively determined by
the classical Hamiltonean. The measure is invariant under the action of the 
Euclidean group in $\R^d$ and provides simultaneously a \rep\ of the
Weyl relations and of the action of the  Poincar\'e group. In particular, one obtains
a  natural quantization of the dynamics \cite{BSZ,RS2}. 

It is well known that the \reps\ associated with free fields with different values of the mass
are not unitarily  equivalent. This nonequivalence reflects the fact that the  measures 
corresponding to distinct masses
are supported in different  -- in fact  disjoint   
-- sets of $\S'(\R^d)$. In Section \ref{clivre2} we studied the differences between the supports 
of the  measures for distinct  masses: we singled out  properties which are specific
of the  ``typical  
(quantum) field'' for a given value of the mass. We have therefore looked for a 
characterization  as refined as possible of the set of elements of $\S'(\R^d)$ that effectively
contributes to a given free field measure. Since the support of the measure replaces
the classical  configuration space in the \reps\ of the kinematical algebra,
this analysis  corresponds to a characterization of the  ``quantum configuration space'' 
for a given value of the free field mass.
For this purpose, we considered a  sequence of stochastic variables that test
the behaviour of the field for large distances.  
%Os resultados obtidos com estas vari\'aveis \cite{MTV} s\~ao 
%fisicamente mais transparentes do que os anteriormente 
%dispon\'\i veis \cite{CL,RR,Ro,Hb,K}. 
Finally, let us mention that the study of the support of the free field measures
is important  not only for a good understanding of these models, but also
from the broader perspective of the construction of quantum theories with interactions.
In the usual procedure, the introduction of interactions in a free model
is constrained by the properties of the 
``typical free fields'': the  local (distributional) behaviour leads to the well known
 ultraviolet divergences, whereas  the long range behaviour is related to
 infrared divergences \cite{S,RR,GJ,Ri,RS2,RS3}.

%\medskip
%\section*{Conflict of Interests}
%The author declares that there is no conflict of interest regarding the publication of this paper.

\newpage

%%%%%%%%%%%%%% Bibliography %%%%%%%%%%%%%%%%%%%%%%%

%\section*{References}
%\bigskip

%%%%%%%%%%%%%%%%%%%%%%%%%%%%%%%%%%

%%%%%%%%%%%%%%%%%%%%%%%%%%%%%%%%%

\end{document}